\documentstyle[aps, psfig]{revtex}
\newcommand{\be}{\begin{equation}}
\newcommand{\ee}{\end{equation}}
\newcommand{\ba}{\begin{eqnarray}}
\newcommand{\ea}{\end{eqnarray}}
\newcommand{\bm}{\bibitem}
\newcommand{\om}{\omega}
\newcommand{\lm}{\lambda}
\newcommand{\de}{\delta}
\newcommand{\De}{\Delta}
\newcommand{\Th}{\Theta}
\newcommand{\mpi}{4 m_{\pi}^2}
\newcommand{\calt}{{\cal T}^{ab}_{\mu\nu}}
\newcommand{\mmn}{M^{ab}_{\mu\nu}}
\newcommand{\mnm}{M^{ba}_{\nu\mu}}
\newcommand{\nmn}{N_{\mu\nu}}
\newcommand{\lmn}{L_{\mu\nu}}    
\newcommand{\th}{\tanh(\beta q_0/ 2)}
\newcommand{\vq}{|\vec{q}|}
\newcommand{\la}{\langle}
\newcommand{\ra}{\rangle}
\newcommand{\ole}{\overline}
\newcommand{\mrs}{m_{\rho}^2}
\newcommand{\frs}{F_{\rho}^2}
\newcommand{\ms}{M^2}
\newcommand{\qz}{q_{0}}

\begin{document}

\setcounter{page}{1}

\begin{center}
{\large{\bf QCD sum rules at finite temperature}}\\
\vspace{1cm}
S. Mallik and Krishnendu Mukherjee\\

\vspace{0.5cm}
Saha Institute of Nuclear Physics,
1/AF, Bidhannagar, Calcutta-700064, India

\end{center}

\vspace{1.5cm}

\noindent{\bf Abstract}
We derive thermal QCD sum rules for the correlation function of two vector
currents in the rho-meson channel. It takes into account the leading
non-perturbative corrections from the additional operators, 
which appear due to the breakdown of Lorentz invariance at
finite temperature. The mixing
of the new operators has a drastic effect on their coefficients. The thermal
average of all the operators can be related to that of the quark condensate
and the energy density. The sum rules then yield the temperature dependence
of the parameters of the $\rho$-meson, namely its mass and coupling to the
vector current. Our result is that these parameters are practically
independent of temperature at least up to a temperature of $125$ MeV.
 
\vspace{1.5cm}

\section{Introduction} 
\setcounter{equation}{0}
The QCD sum rules \cite{SVZ}, proposed about two decades ago,
prove remarkably 
successful in addressing the non-perturbative problems of hadron 
phenomenology. In this approach one considers
the product of two local operators,
like the currents of the QCD theory. A sum rule is obtained by equating 
the dispersion relation for its vacuum matrix element at sufficiently  
large space-like momenta to the corresponding Wilson operator product 
expansion \cite{Wilson}. The higher dimension operators
present in the expansion give rise to 
the non-perturbative, power corrections. The idea of extending 
these sum rules to finite temperature by replacing the vacuum matrix
element with the thermal average
naturally suggests itself.

The original work establishing the thermal QCD sum rules is that of
Bochkarev and Shaposhnikov \cite{Shap}. They recognised the importance of
the (low energy)
continuum states in the spectral function to account for the
effects of the medium. On the basis of these sum rules 
they discussed the temperature dependence of the resonance parameters 
and the existence of phase transition. However, there arise 
additional operators
in the Wilson expansion at finite temperature \cite{Shur},
 which were not correctly 
incorporated in their sum rules: In effect, they calculated these new 
operator contributions perturbatively, which cannot be justified,
particularly at low temperature.  

The additional operators 
arise because of the breakdown of Lorentz invariance
at finite temperature by the choice of the thermal rest frame, where
matter is at rest at a definite temperature \cite{Weldon82}. The residual 
O(3) invariance naturally brings in additional operators. The expected 
behaviour of the thermal averages of these Lorentz non-invariant (new)
operators is 
somewhat opposite to those of the Lorentz invariant (old) ones: 
While the old
operators start with non-zero values at zero temperature
and decrease in magnitude with the rise of temperature, 
the new ones, on the other hand, are zero
at zero temperature but grow rapidly with temperature. The importance 
of including these new operators in the thermal sum rules,
particularly at not too low a temperature, is thus clear.

Although a number of works on thermal QCD sum rules exist by
now, these are flawed with respect to the new operators: Either some of
these are missed \cite{Hatsuda} or their mixing, which
changes their coefficients drastically, is not taken into account
\cite{Eletsky}. 
In a recent work \cite{Mallik97} we applied
a simple, configuration space method \cite{Fritzsch},\cite{Hub}
to evaluate the Wilson coefficients of these new operators (up to dimension 
four) which appear at finite temperature in the short distance expansion of
the product of two quark bilinear operators. Here we make use of this 
result to write and evaluate the
thermal QCD sum rules, incorporating correctly the contributions from all
the dimension four operators. 

We consider the correlation function of the time ordered $(T)$ product of two
vector currents in the $ \rho$-meson channel. The use of the $T$-product, rather
than the retarded (or advanced) product, is a little complicated
in writing down the spectral representation but has the advantage
in perturbative calculations, for which we can apply the
conventional formalism.
Throughout this work we shall employ the real time formulation of
the thermal field theory \cite{Niemi} , which requires in general not only
the physical fields but also the accompanying 'ghost' fields. Since, 
however, we work to lowest order in perturbation expansion, ghost 
fields do not show up.

It is convenient to write difference sum rules by subtracting the
vacuum sum rules from their finite temperature counterparts. For such
sum rules the absorptive parts are expected to be saturated with the
$\rho$-meson pole and the $\pi\pi$-continuum \cite{Shap}, 
whose contributions we derive
here for completeness. The thermal averages of the different
operators reduce essentially to that of the quark condensate and the energy
densities of quarks and gluons \cite{Leutwyler92}.
Chiral perturbation theory \cite{Gasser} has been used to calculate the
temperature dependence of the quark condensate
\cite{Gerber}. The difficulty with the energy densities
is that while one of them is the total energy density, which can be obtained
from a knowledge of the hadronic spectrum at low temperature, the other is
an unphysical combination of the two densities. We need an additional
input to relate this latter
combination to the total energy density.

Once the power corrections are known, the difference sum
rules give the temperature dependence
of the $\rho$-meson parameters, namely its mass and its coupling with the
vector current. With our simple saturation scheme, the sum rules can be used
up to a temperature of about 125 MeV. The numerical evaluation shows that
these parameters have negligible dependence on temperature.

In sec.II we write the kinematic decomposition for the thermal
correlation function of two vector currents, derive the Landau
representation \cite{Landau}
for the time ordered product and state the results of 
operator product expansion to derive finally the form of the thermal QCD sum
rules. In sec.III we obtain the contributions of $\rho$ and $\pi\pi$  
intermediate states to the spectral function. 
In sec.IV we collect the information on thermal average of the operators
present in the sum rules and evaluate the difference sum rules. In sec. V we
discuss how to extend the sum rules to
higher temperature. 
In the Appendix we give the details of the evaluation of a limit  
stated earlier\cite{Shap}.

\section{Sum rules}
\setcounter{equation}{0}
\renewcommand{\theequation}{2.\arabic{equation}}

We derive the QCD sum rules for 
the thermal average of the time ordered ($T$) product of
two currents,
\be
\calt (q) = i \int{d^4x} e^{iq\cdot x}
\left \langle T\left(V^a_{\mu}(x)
V^b_{\nu}({\rm o})\right) \right \rangle.
\ee
Here $V_{\mu}^a (x)$ is the vector current (in the $\rho$ meson channel)
in QCD theory,
\be
V^a_{\mu}(x) =\bar{q}(x) \gamma_{\mu}(\tau^a /2) q(x), 
\ee
$q(x)$ being the field of the $u$ and $d$ quark doublet and $\tau^a$ the
Pauli matrices. The thermal average of an operator $O$ is denoted by 
$ \la O \ra$,
\[\langle O\rangle =Tr e^{-\beta H} O/Tr e^{-\beta H} ,\]
where H is the QCD Hamiltonian, $\beta$ is
the inverse of the temperature $T$ (coinciding with the time ordering 
symbol !) and $Tr$ denotes the trace over any complete
set of states. 

\subsection{ Kinematics}

At finite temperature Lorentz invariance is broken by the choice of a
preferred frame of reference, viz, the matter rest frame where temperature
is defined. But  the book-keeping with indices
becomes simpler if we restore it formally by introducing the four-velocity
$u_{\mu}$ of the matter \cite{Weldon82}. 
(In the matter rest frame 
$u_{\mu}=(1,0,0,0)$.)
The time and space components of $q_{\mu}$ are then raised to the 
Lorentz scalars, $\om = u\cdot q$ and $\bar{q}=\sqrt{\om^2-q^2}$.
 We shall, however, return to the matter
rest frame while doing all actual computations.

In such a Lorentz invariant framework, there are two independent conserved
kinematic covariants \cite{Kapusta}, which we choose as

\[P_{\mu\nu}=-\eta_{\mu\nu} +{q_{\mu}q_{\nu}\over q^2} -{q^2\over \bar{q}^2}
\tilde{u}_{\mu} \tilde{u}_{\nu}, \qquad Q_{\mu\nu}={q^4\over \bar{q}^2}
\tilde{u}_{\mu}\tilde{u}_{\nu},\]
where $\tilde{u}_{\mu} = u_{\mu} -\om q_{\mu}/ q^2$. The invariant
decomposition of ${\cal T}^{ab}_{\mu\nu}$ is then given by
\be
{\calt} (q) =\de^{ab} (Q_{\mu\nu} T_l + P_{\mu\nu} T_t),
\ee
where the invariant amplitudes $T_{l,t}$ are functions of $q^2$ and $\om$.

Notice that the kinematic covariants $P_{\mu\nu}$ and $Q_{\mu\nu}$ are free
from singularities at $q^2=0$ (and also at $ \bar{q}^2=0$ ). This is convenient
at finite temperature as there are dynamical singularities extending up to 
$q^2=0$. With this choice of kinematic covariants, the dynamical
singularities reside only in the invariant amplitudes.

To extract the invariant amplitudes from $\calt$, it is converient to form
the scalars $\Pi_{1,2}$,
\[ \de^{ab} \Pi_1 (q) ={\cal T}_{\mu}^{\mu ab} (q), \quad
\de^{ab} \Pi_2 (q) = u^{\mu} \calt (q) u^{\nu} .\]
Then the invariant amplitudes are given by 
\be
T_l ={1\over \bar{q}^2} \Pi_2,\qquad T_t =-{1\over 2} \left ( \Pi_1
+{q^2\over \bar{q}^2} \Pi_2 \right ).
\ee

In the limit $\vec{q} =0$, the amplitudes $T_l $ and $T_t$ are related. To
see this we write the spatial components of ${\cal T }_{\mu\nu}$ as
\[{\cal T}_{ij}^{ab}(q) 
= \de^{ab}[\de_{ij} T_t -\hat{q}_i\hat{ q}_j (T_t -q^2_0 T_l)],\]
where $\hat{q}_i$ is the $i$th component of the unit vector along $\vec{q}$. As
$\vq\rightarrow 0$, there cannot be any dependence on $\vq$, getting
\be
T_t(q_0, \vq =0)=q_o^2 T_l(q_0,\vq=0).
\ee

\subsection{ Spectral representation}

Let us obtain the spectral representation for the correlation function in
$q_0$ at fixed $\vq$. First, evaluate the trace over a complete set
of eigenstates of four-momentum, when it becomes a sum over forward 
amplitudes weighted by the corresponding Boltzmann factors. Then insert 
the same set of complete states between the currents to 
extract its $x$- dependence which is then integrated out. Introducing
a $\de$- function in $q_0$, the result can be written as
\be
\calt (q) = {1\over{2\pi}} \int^{+\infty}_{-\infty}dq^{\prime}_0 \left(
{{\mmn (q^{\prime}_0, \vq)}\over{q^{\prime}_0 - q_0 - i\epsilon}} -     
{{\mnm (q^{\prime}_0, \vq)}\over{q^{\prime}_0 - q_0 + i\epsilon}}\right),
\ee    
where the expression for $\mmn$ with the double sum over states 
may be converted back to the form, 
\be
\mmn (q) =  \int d^4 x e^{iqx} \langle V^a_{\mu}(x)
V^b_{\nu}({\rm o})\rangle.
\ee 
Using the double sum representation,
it is easy to show that
\be
\mnm(q_0 , \vq) = e^{-\beta q_0}\mmn(q_0 , \vq).
\ee 

The opposite sign of $i\epsilon$ in the two terms in eqn.(2.6) is 
typical of $T$- products. As a result the imaginary part of $\calt$ 
is given by the
sum, ${1\over2}(\mmn + \mnm)$, while the principal value integral 
contains the difference, ${1\over 2}(\mmn - \mnm)$.  
But the two are related by (2.8),
\be
\mmn - \mnm = \left(\mmn + \mnm\right) \th .
\ee
We thus get the Landau representation for the $T$ product at finite 
temperature \cite{Landau},
\be
\calt (q_0,\vq) = i Im\calt (q_0,\vq) + P
\int^{+\infty}_{-\infty}dq^{\prime}_0 {\nmn(q^{\prime}_0,\vq)\over
{q^{\prime}_0 - q_0}},
\ee
where, for brevity, we write
\be
\nmn^{ab} (q)\equiv
\de^{ab} \nmn (q) =\pi^{-1} Im \calt (q)\th 
\ee

Expressing $q_\mu ^{\prime} =q_{\mu} + (q_0^{\prime} -q_0)u_{\mu}$, we
recover the representation for the invariant amplitudes $T_{l,t}$.
Further, using the symmetry properties, $Im T_{l,t} (-q_{\mu})
 = Im T_{l,t} (q_{\mu})$ and 
going over to imaginary values of $q_0$ $(q^2_0=-Q^2_0,\; Q^2_0>0)$,
for which $\nmn$ vanishes, these become, 
\be
T_{l,t}(q_0^2,\vq) =
\int^{\infty}_0dq^{\prime^2}_0 {N_{l,t}(q^{\prime}_0,\vq)\over
{q^{\prime^2}_0 + Q_0^2}}.
\ee
It may actually require subtractions, but it does not affect 
the Borel transformed sum rules we shall write below.  

\subsection{ Operator product expansion}

The contributions of operators of dimension four to $T_{l,t}$ 
are obtained from the short distance expansion and improved upon by the
renormalisation group equation
in a previous work \cite{Mallik97}. Including the unit operator
(the perturbative contribution), $T_l$ is given for large Euclidean 
momenta by $(Q^2=-q^2)$,
\be
T_l = -{1\over 8\pi^2}\ln({Q^2\over\mu^2})
 + {1\over Q^4}\left ( \hat{m}\la \bar{u}u\ra + {\la G^2\ra\over 24}
 + {4\over{16+3n_f}}\{ \la \Th \ra +a(Q^2)\la 16\Th^f/3-\Th^g\ra\} \right ).
\ee
$T_t$ is also given by the same 
expression, except for an overall factor of $-Q^2$
and a factor of $(1-2\bar{q}^2/ Q^2)$ multiplying the term with $\Th$'s.
Here $\hat{m}$ is the degenerate mass of the $u$ and the $d$ quarks and
${1\over 2}\la\bar{q}q\ra =\la\bar{u}u\ra =\la\bar{d}d\ra$. 
The operator $G^2$ is quadratic in
the gauge field strength $G^a_{\mu\nu}, (a=1,\cdots ,8), 
\; G^2=(\alpha_s/{\pi})
G^a_{\mu\nu} G^{a\mu\nu}$, with
$\alpha_s =g^2/{4\pi}$, where  
$g$  is the gauge coupling constant.  
Along with these two operators already
contributing to zero temperature, we now have the linear combinations of
two new ones, ${\Th}^{f,g} \equiv u^{\mu} {\Th}_{\mu\nu}^{f,g} u^\nu$, where
${\Th}_{\mu\nu}^{f,g}$ are the energy-momentum tensors for quarks 
of a single flavour and of
gluons. The corresponding components for the total tensor is $\Th=
n_f\Th^f +\Th^g$, where $n_f$ is the number of effective quark
flavours. The logarithmic $Q^2$ dependence of $a(Q^2)$ arise due to the
anomalous dimension $d$ of the operator $(16\Th^f/3-\Th^g)$,
\[a(-q^2)=\left ( {\alpha_s(\mu^2)\over \alpha_s(-q^2)} \right ) ^{-d/2b},
\qquad
d={4\over 3}({16\over 3} + n_f), \qquad b =11- {2\over 3}n_f,\]
where $\mu (\simeq$ 1 GeV) is the scale at which  all 
renormalisations are carried out.

Note the mixing of the operators $\Th^f$ and $\Th^g$ in (2.13) under the
renormalisation group, which is well-known in the context of deep inelastic
scattering \cite{Peskin}.
In the operator product expansion of two quark currents, the
Wilson coefficients of $\Th^f$ and $\Th^g$ are, to leading order, of zeroth
and first order in $\alpha_s$, arising from Born and one-loop graphs
respectively. However, due to operator mixing, the coefficients are
drastically changed in that both $\Th$ and $(16\Th^f/3-\Th^g)$ 
have coefficients
with leading terms of zeroth order in $\alpha_s$. In (2.13) we retain only these
leading terms.

\subsection{ Sum rules}

We now equate the spectral representation and the result of
operator product expansion for the amplitudes $T_l$ and $T_t$
at sufficiently high $Q_0^2$.  Taking Borel transform we arrive at the
thermal QCD sum rules \cite{Shap}. For $T_l$ we get 
\be
{1\over M^2}\int^{\infty}_0 dq^2_0  
e^{-q^2_0/ M^2} N_l (\qz,\vq) 
= e^{-\vq^2/ M^2}\left({1\over 8\pi^2} + {\la O \ra \over M^4}\right),
\ee   
where $\la O \ra$ is the non-perturbative contribution of higher dimension 
operators,
\be
\la O \ra = \hat{m} \la \bar{u}u \ra + {\la G^2 \ra \over 24}+
{4\over{16+3n_f}}\{\la \Th \ra + a(M^2)  \la 16\Th^f/3 -\Th^g \ra \}.
\ee
and a similar one for $T_t$. Each is a two parameter sum rule, dependent not
only on $T$ but also on $\vq$. 

In the thermal rest frame the thermal average of the new operators are
energy densities, which increase rapidly with temperature. 
Earlier works on thermal QCD sum rules were flawed, as
these contributions were not properly included.

\section{Absorptive parts}
\setcounter{equation}{0}
\renewcommand{\theequation}{3.\arabic{equation}}

We work below the critical temperature, where hadrons constitute the
physical spectrum. As with the vacuum sum rules, the dominant 
contribution to the spectral function is given by the $\rho$-meson.
We also calculate the contribution of the non-resonant $\pi\pi$-continuum.
The question of other significant contributions will be discussed in Sec. IV.

\subsection{ $\rho$-pole}
The coupling of the vector current to the $\rho$-meson is given by
\be
<0|V_{\mu}^a|\rho^b> = \de^{ab} F_{\rho} m_{\rho} \epsilon_{\mu},
\ee
where $\epsilon_{\mu}$ is the polarisation vector of $\rho$ of mass
$m_{\rho}$. The experimental value of $F_{\rho}$ as measured by the
electronic decay rate of $\rho^0 $ \cite{Particles} is $F_{\rho} = 153.5$
MeV. 

A simple way to calculate the absorptive part 
of the $\rho$-pole diagram is to note
the field-current identity of $V_{\mu}^a (x)$ with the rho-meson field
$\rho_{\mu}^a$,
\be
V^a_{\mu}(x) = m_{\rho}F_{\rho}\rho^a_{\mu}(x),
\ee
which reproduces (3.1). Then the $\rho$-meson 
contribution is given essentially by its thermal propagator,
\ba
\calt(q) &=& i m^2_{\rho}F^2_{\rho}\int d^4x e^{iqx}
\la \rho^a_{\mu}(x)\rho^b_{\nu}({\rm o}) \ra\nonumber\\
&=& i \de^{ab} m^2_{\rho}F^2_{\rho}\left (-\eta_{\mu\nu}
+{q_{\mu} q_{\nu}\over \mrs} \right )\De^{\rho}_{11}(q), 
\ea
where $\De^{\rho}_{11}(q)$ is the $11$- component of a scalar field
propagator with mass $m_{\rho}$, 
\be
\De^{\rho}_{11}(q) = {i\over {q^2 - m^2_{\rho} + i\epsilon}} + 
2\pi n(\om_q)\de(q^2-m_{\rho}^2), 
\ee
and $n(\om_q)$ is the Bose distribution function, $n(\om_q)=
(e^{\beta \om_q}-1)^{-1}$, $\om_q=\sqrt{{\vec q}^2+m^2_{\rho}}$. We then
get 
\be
\left(\begin{array}{c}N_l\\N_t\end{array}\right) =  
\left(\begin{array}{c}1\\m_{\rho}^2\end{array}\right) \frs \de \{q_0^2
-(\vq^2 +\mrs)\}
\ee  
Loop corrections at finite temperature will make $m_{\rho}$ and $F_{\rho}$
 temperature
dependent, modifying them, in general, differently in the longitudinal and
transverse amplitudes. These modifications may be obtained by calculating the
appropriate loop graphs. Here we shall find them by evaluating our sum
rules.

\subsection{$\pi\pi$-continuum}
The $\pi\pi$-contribution to the amplitudes describes the interaction of the
current with the particles in the medium, which are predominantly pions.
Chiral perturbation theory \cite{Gasser}
gives the contribution of the pion field $\phi^a
(x)$ to the vector current, which, to lowest order, is
\[ V^a_{\mu} (x) =\epsilon ^{abc} \phi^b (x) \partial_{\mu} \phi^c (x) \]
Then the pions contribute to the correlation function as
\be
\calt (q)= i \de^{ab}\int{d^4k\over (2\pi)^4}(2k-q)_{\mu}(2k-q)_{\nu}
\De^{\pi}_{11}(k)\De^{\pi}_{11}(k-q),
\ee
where $\De^{\pi}_{11}$ is again the $11$- component of the scalar 
propagator (3.4) but with mass $m_{\pi}$. Its imaginary part 
can be obtained by the cutting
rules at finite temperature \cite{Kobes}. Here we  obtain it directly for
this simple amplitude. We express $\De^{\pi}_{11}$ as
\be
\De^{\pi}_{11}(k) = \{1 + n(\om_k)\}D(k) + n(\om_k)D^*(k),
\ee
where $D(k)$ is the zero temperature propagator, $D(k)=i/(k^2 - m^2_{\pi}
 + i\epsilon)$, and carry out the $k_o$- integration \cite{foot2}.
The imaginary
part may then be read off as
\be
Im\calt (q)= \de^{ab} \{\lmn(q) + \lmn(-q)\},
\ee
where
\ba
\lmn(q) &=& \pi \int{d^3k\over (2\pi)^3} {{(2k-q)_{\mu}(2k-q)_{\nu}}\over
{4\om_1\om_2}}\nonumber\\
& & [\{(1 + n_1)(1 + n_2) + n_1n_2\}\de(q_0 - \om_1 - \om_2)\nonumber\\
& &~~+ \{(1 + n_1)n_2 + (1 + n_2)n_1\}\de(q_0 - \om_1 + \om_2)]. 
\ea
Here $n_1\equiv n(\om_1)$, $n_2\equiv n(\om_2)$ with $\om_1=\sqrt
{{\vec k}^2+m^2_{\pi}}$, $\om_2=\sqrt{({\vec k}-{\vec q})^2+m^2_{\pi}}$.
The time component of $k_{\mu}$ in the tensor structure is 
understood to be given by $k_0=\om_1$.
 
With the help of the
$\delta$-functions we can rerwite (3.9)  to get $N_{\mu\nu}$
defined by (2.11) for $q_0>0$,    
\ba
\nmn(q) &=&  \int{d^3k\over (2\pi)^3} {{(2k-q)_{\mu}(2k-q)_{\nu}}\over
{4\om_1\om_2}}\nonumber\\
& & [(1 + n_1 + n_2)\de(q_0 - \om_1 - \om_2) + (n_2 - n_1)
\de(q_0 - \om_1 + \om_2)].
\ea
In this form the 
factors involving the density distributions can be interpreted in
terms of pion absorption from and emission into 
the medium \cite{Weldon83}. Since the
first and the second $\de$- functions in (3.10) contribute to
time-like ($q^2\geq \mpi$) and space-like ($q^2<0$) regions respectively, 
we write it as
\ba
\nmn(q) &=& \int{d^3k\over {(2\pi)^3 2\om_1}} {(2k-q)_{\mu}(2k-q)_{\nu}} 
\nonumber\\
& & ~[(1 +n_1 + n_2)\theta(q^2 - \mpi) + (n_2 - n_1)\theta(-q^2)]
\de((q_0 - \om_1)^2 - \om^2_2).
\ea
Thus in addition to the usual cut, $4m_{\pi}^2 +\vq^2 \leq q_0^2\leq
\infty$, the amplitude at finite temperature acquires 
a new short cut, $0\leq q_0^2 \leq \vq^2  \;$ \cite{Weldon83},
\cite{Shap}.

The angular integration is carried out using the $\de$- function, 
when the constraint $|\cos\theta_{{\vec q},{\vec k}}|\leq 1$ results in a 
$\theta$- function, $\theta[-q^2(\om_1-\om_+)(\om_1-\om_-)]$, where
\be
\om_{\pm}={1\over 2}(q_0\pm \vq v),\qquad 
v(q^2)=\sqrt {1-{\mpi/ q^2}}.
\ee
We thus get
\ba
\nmn(q)&=&{1\over 2|{\vec q}|}\int^{\om_+}_{\om_-}{d\om_1\over (2\pi)^2}
{(2k-q)_{\mu}(2k-q)_{\nu}}(1 + n_1 + n_2)\theta(q^2 - \mpi)\nonumber\\
& &+ {1\over 2|{\vec q}|}\int^{\infty}_{\om_+}{d\om_1\over (2\pi)^2}
{(2k-q)_{\mu}(2k-q)_{\nu}}(n_2 - n_1)\theta(-q^2).
\ea

It is now simple to extract the absorptive parts of the invariant
amplitudes by using (2.4). Changing the variable $\om_1$ to $x$
given by $\om_1={1\over 2}(q_0+\vq x)$, we get   
\be
\left(\begin{array}{c}N_l^+\\N_t^+\end{array}\right) =  
{{v^3}\over 48\pi^2}\left(\begin{array}{c}1\\q^2\end{array}\right) +    
\left(\begin{array}{c}{\bar N}_l^+\\{\bar N}_t^+\end{array}\right),
~~~~~ {\rm for}~~~ q^2>\mpi
\ee
with
\be
\left(\begin{array}{c}{\bar N}_l^+\\{\bar N}_t^+\end{array}\right) =
{1\over 32\pi^2}\int^{v}_{-v} dx
\left(\begin{array}{c}{2x^2}\\q^2(v^2-x^2)
\end{array}\right)n((\vq x+q_0)/2)
\ee
and
\ba
\left(\begin{array}{c}N_l^-\\N_t^-\end{array}\right) &=&
{1\over 64\pi^2}\int^{\infty}_v dx
\left(\begin{array}{c}{2x^2}\\q^2(v^2-x^2)
\end{array}\right)\nonumber\\
& & \{n((\vq x-q_0)/2)-n((\vq x+q_0)/2)\},
~~~~{\rm for}~~~ q^2\leq 0
\ea  
The superscripts $(\pm)$ on $N$ denote time-like and space-like
$q_\mu$ respectively, 
where they are non-vanishing. The first term on the right of (3.14)
arising from the unity in the factor $(1+n_1+n_2)$ in (3.13) 
is the zero temperature 
contribution of the $\pi\pi$ state. Evaluated here in a non-covariant way
it, of course, agrees with the covariant evaluation of the Feynman amplitude
(3.6)
with $\Delta^{\pi}_{11}(k)$ replaced by the vacuum propagator 
$D(k)$ \cite{Gasser}. 

\subsection{Explicit sum rules}
 
Let us now write explicitly the sum rule (2.14) for $T_l$.
Saturating $N_l$ with
the above contributions it becomes,
\ba
& & F_\rho^2(T) e^{-m_\rho^2(T)/M^2} + 
{1\over {48 \pi^2}} \int_{\mpi}^\infty dq_0^2 e^{-{q_0^2/M^2}}
v^3(q_0^2 +\vq^2)\nonumber\\
&+& e^{\vq^2/M^2}\left( \int_{\mpi+{\vq}^2}^\infty dq_0^2 
e^{-{q_0^2/M^2}}{\bar N_l}^+(q_0,\vq) +
\int_0^{{\vq}^2}dq_0^2 e^{-{q_0^2/M^2}} 
N_l^-(q_0,\vq)\right)\nonumber\\
&=&  {M^2\over 8\pi^2} + {\la O \ra\over M^2} 
\ea
As the temperature goes to zero, the two terms in bracket go to zero and
the thermal average of the operators on the right become the
vacuum expectation values, recovering the familiar vacuum sum rule.
The integral on the left is the
non-resonant $2\pi$ contribution and is small compared to
the resonance contribution.

As $\vq\rightarrow 0$, the sum rule (3.17) simplifies
considerably. The limit for the second integral in bracket is given in ref.
\cite{Shap}. Here we derive it in the Appendix. The sum rules for $T_l$ and
$T_t$ then become,

\be
  F_\rho^2(T) e^{-{m_\rho^2(T)/M^2}} + I_0 (M^2) +I_T (M^2)=
  {M^2\over 8\pi^2} + {\la O \ra\over M^2},
\ee
and
\be
m_{\rho}^2 (T) F_\rho^2(T) e^{-{m_\rho^2(T)/M^2}} + J_0 (M^2) +J_T (M^2)=
  {M^4\over 8\pi^2} -\la O \ra,
\ee
where
\ba
I_0 (M^2)&=&{1\over 48\pi^2} \int ^{\infty}_{4m_{\pi}^2} ds
e^{-s/M^2} v^3, \nonumber\\
J_0 (M^2)&=&{1\over 48\pi^2} \int ^{\infty}_{4m_{\pi}^2} ds s
e^{-s/M^2} v^3,  \nonumber\\
I_T (M^2)&=&{1\over 24\pi^2} \int ^{\infty}_{4m_{\pi}^2} ds
\{e^{-s/M^2} v^3 + v (3-v^2)/2\} n(\sqrt{s}/2),  \nonumber\\
J_T (M^2)&=&{1\over 24\pi^2} \int ^{\infty}_{4m_{\pi}^2} ds s
e^{-s/M^2} v^3 n(\sqrt{s}/2),
\ea 
with $v\equiv v(s)=\sqrt{1-4m_{\pi}^2/s}$.
Observe that the sum rules (3.18-19) are not independent, in 
agreement with the relation (2.5): the second one 
is obtained by differentiating the first
with respect to $1/M^2$. 

\section{Evaluation of sum rules}
\setcounter{equation}{0}
\renewcommand{\theequation}{4.\arabic{equation}}
In the above sum rules we have 
approximated the absorptive parts of the amplitudes
by the $\rho$-meson pole and the $2\pi$ continuum, while we retain only the
contributions of the unit operator (the perturbative result) and of all the
operators of dimension four in the operator product expansion. To check this
saturation scheme, let us compare the zero temperature limit of our sum
rules with the corresponding vacuum sum rules \cite{SVZ}. The
latter include, in addition, the rather large contribution from
the high energy 
continuum beyond $1.5$ GeV, as indicated by the experimental data, as well
as the contribution of a quark operator of dimension six, which is also
large because of its origin in Born (rather than loop) diagram. Since we do
not include any of these contributions, we cannot expect the sum rules as
written above to be well saturated.

Rather than incorporate these contributions, 
we isolate the thermal effects 
by considering the difference
sum rules, obtained by subtracting out the vacuum sum rules from
the corresponding finite temperature ones. Then the contribution to the
absorptive parts beyond 1.5 GeV, being temperature independent, cancels out
in the difference. Thus it is the temperature dependent contributions of the
$\rho$-meson and of the $2\pi$ continuum, which should dominate the
absorptive parts of these sum rules. Also the dimension
six quark operator, $O_6$ say, contributes an amount
$\sim  \la O_6\ra -\la 0|O_6|0 \ra$, which is
insignificant in the immediate neighbourhood of $T=0$ and as the temperature
rises, this contribution is overwhelmed by that of the two-gluon and other
Lorentz non-invariant 
operators, as our estimates below for these operators show.

The difference sum rules for the two invariant amplitudes 
allow us to calculate the temperature dependence
of the $\rho$-meson parameters,
\ba
\Delta m_{\rho} (T)&\equiv& m_{\rho}(T)-m_{\rho} \nonumber \\
 & =&{ m_{\rho} e^{\mrs /\ms}\over 2\frs}\left \{I_T -{J_T\over \mrs}
-\left ( {1\over \mrs} +{1\over  \ms} \right ) \ole {\la O\ra} \right \},
\ea
\ba
\Delta F_{\rho} (T) & \equiv & F_{\rho}(T) -F_{\rho} \nonumber \\
&=& - {e^{\mrs/\ms}\over {2F_{\rho}}} \left \{ {J_T\over \ms}
+\left (1-{\mrs\over \ms} \right )I_T +{\mrs\over M^4} \ole{\la O \ra}
\right \},
\ea
with
\be
\ole{\la O \ra} = \hat{m} \ole{\la \bar{u}u \ra} + {\ole{\la G^2 \ra}
 \over 24}+
{4\over{16+3n_f}}\{\la \Th \ra + a(M^2)  \la 16\Th^f/3 -\Th^g \ra \},
\ee
where the bar on the operators indicates subtraction of their vacuum
expectation values. Here we insert the experimental values
for $m_{\rho}$ and $F_{\rho}$, as these are
well reproduced by the vacuum sum rules.

We now collect information on the operator contributions. The vacuum
expectation value of the chiral condensate $\la 0| \bar{q} q |0 \ra$ is known
from the PCAC relation of Gell-Mann, Oakes and Renner,
\[ 2\hat{m} \la 0|\bar{u}u |0\ra = - F_{\pi}^2 m_{\pi}^2, \]
where $\hat{m}={1\over 2} (m_u +m_d) $ and the pion decay constant 
$F_{\pi}$, defined by,
\[ \la 0| A^a_{\mu} |\pi ^b (q)\ra =iq_{\mu} \de^{ab} F_{\pi} ,\]
has the value $F_{\pi}=93.3$ MeV. Taking $\hat{m} =7$ MeV \cite{Leutwyler96},
 we get 
$  \la 0|\bar{u}u |0\ra = - (225MeV)^3 $. The vacuum expectation value of
the two-gluon operator, as determined from the QCD sum rules \cite{SVZ}, is
$\la 0| G^2|0\ra = (330 MeV)^4$.

The operator $G^2$ is related to the trace of the energy momentum tensor
$\Th_{\mu\nu}$ 
by the trace anomaly. Normalising it to zero vacuum expectation value
and taking thermal average, it gives
\ba
\ole{\la G^2\ra} &\equiv & \la G^2\ra -\la 0|G^2|0 \ra \nonumber \\
&=&-{8\over 9} \left (\la \Th^{\mu}_{\mu}\ra -\sum_q m_q 
\ole{ \la \bar{u}u \ra } \right ).
\ea
The trace at finite temperature is given by 
$ \la \Th_{\mu}^{\mu} \ra =\la \Th \ra -3P, $
where $\la \Th \ra$ is the energy density and $ P$ the pressure.

The temperature dependence of both  $ \la \bar{u}u \ra$ and  $\la \Th \ra$
have been calculated in chiral perturbation theory \cite{Gerber}.
 Corrections due to nonzero quark masses as well as the contributions  
of the massive states ($K,\eta,\rho,\cdots$) have also been incorporated.
However, as the authors point out, the validity of the perturbation
theory and the use of dilute gas approximation  to calculate the
contribution of the massive states restrict these results to within a
temperature of about $150 MeV$. Thus the critical temperature $T_c$ is,
strictly speaking, beyond the range of validity of their calculation. Since,
however, the order parameter $\la \bar{u}u \ra$ 
falls rapidly at the upper end of
this range, one has only to extrapolate it a little further to get
$T_c =190 MeV$.

Besides the total energy density $\la\Th\ra$, there also occurs 
the thermal average, $\la 16\Th^f/3 -\Th^g \ra $. The last one cannot be 
calculated 
without further input, at least in the hadronic phase. Now both naive
counting of the degrees of freedom and empirical study of the pion structure
function \cite{Shur} suggest the quark fraction of the energy density to
be about half of the total. So we assume
 $n_f\la\Th^f \ra=\chi_f \la \Th \ra$,with $\chi_f =.5$, whence $\la
16\Th^f/3
-\Th^g \ra = \{({16\over {3n_f}}+1)\chi_f -1\}\la \Th \ra$.


\begin{figure}
\centerline{\psfig{figure=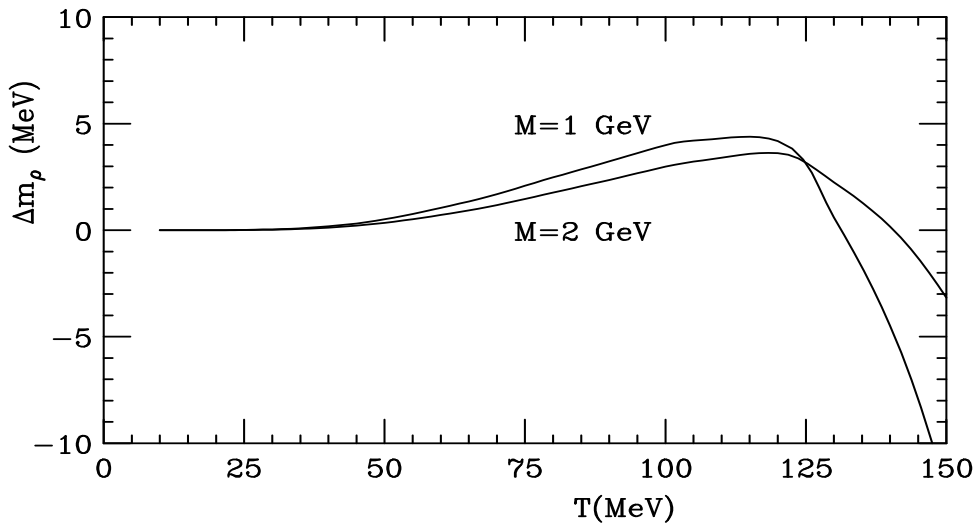,width=13.0cm,height=7.0cm}}
\bigskip
\caption[]{Shift in the rho-meson mass as a function of temperature
for $M^2=1~{\rm GeV}^2$ and $M^2=4~{\rm GeV}^2$.}
\end{figure}

As with the zero temperature sum rules, the results are expected to be
stable in a region in $M^2$, which is
neither too high to make the continuum contribution 
large relative to the resonance contribution nor
too low to emphasize the neglected power corrections of higher order.
Since the high energy continuum contribution gets cancelled in the
difference sum rules, the region of $M^2$ may be extended somewhat at the
upper end. Figs. 1 and 2 show the evaluation for $M^2$ equal to $1 GeV^2$
and $4 GeV^2$. The results for $\De m_{\rho}$ and $\De F_{\rho}$ are 
rather stable for
temperatures up to about 125 MeV. At higher temperatures the results, 
particularly for $\De m_{\rho}$ appear unstable. Closer observation
reveals here a large cancellation between 
the $2\pi$ contribution and the leading power correction we have retained.
Thus the non-leading power correction become important here, whose inclusion
may restore the stability in $M^2$ to higher temperatures.
\begin{figure}
\centerline{\psfig{figure=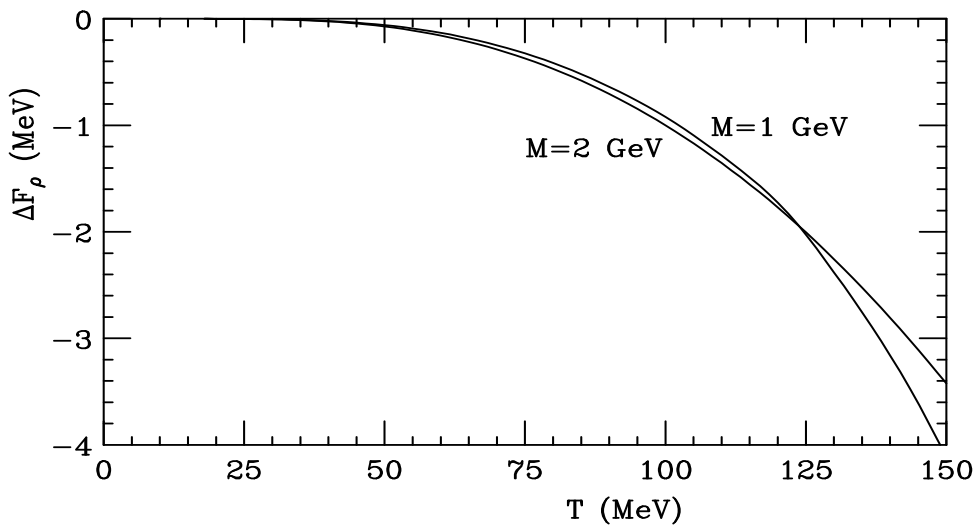,width=13.0cm,height=7.0cm}}
\bigskip
\caption[]{Shift in the coupling of rho-meson with the vector current
as a function of temperature
for $M^2=1~{\rm GeV}^2$ and $M^2=4~{\rm GeV}^2$.}
\end{figure}

\section{Discussions}

In this work we have written the thermal QCD sum rules 
for the two point correlation function of  vector currents in the 
$\rho$-channel,
including the leading power correction 
due to  all the operators of
dimension four. Because of the loss of Lorentz invariance at finite 
temperature, two new (Lorentz noninvariant) 
operators creep up in the Wilson expansion,
 in addition to the two  old (Lorentz invariant) ones, already
existing in the vacuum sum rules. Thus compared to the two numbers, 
$\la 0|\bar{q} q |0 \ra$ and $ \la 0 |G^2 |0 \ra$ in the vacuum sum rules,
we have now four temperature dependent quantities,
$\la \bar{q} q \ra, \la G^2 \ra, \la \Th\ra$ and 
$\la 16 \Th^f /3-\Th^g \ra $
in the thermal sum rules. All these thermal averages can be evaluated 
from chiral perturbation theory, supplemented by contributions from the
massive states. An extra input is needed only for the last unphysical
operator. The sum rules can then be used to 
determine the temperature dependence of the
$\rho$-meson parameters. Our evaluation shows that the mass of 
the $\rho$-meson and its coupling with the
vector current remain practically unaffected by the rise of temperature up
to at least 125 MeV. The absence of the shift in the mass appears to agree
with one set of results obtained in ref\cite{Song}.

Unfortunately the sum rules, as they stand, cannot be extended up to the
critical temperature. One reason contributing to this restriction has to
do with the evaluation of the thermal average of the operators, as we
already discussed in the last section. What restricts it further
is their instability under a change of $M^2$ for temperatures above $125
MeV$.
Even in this restricted temperature range, the numerical evaluation shows
that the new operators are significant. In fact, for temperatures above say
70 MeV, the new contributions overwhelm the old ones in the difference sum
rules. This situation necessitates
reanalysis of all earlier results based on thermal QCD sum rules, where the 
new operators are not properly taken into account.

We now discuss a possible way to extend the sum rules to higher
temperatures.
In the vacuum sum rules the operators of dimension four (and
higher) provide corrections to the leading (perturbative) result. But in the
difference sum rules it is these corrections which become the leading
contribution. One thus expects that by including nonleading contributions
from higher dimension operators along with those of dimension four already
included, the sum rules would be stable against variations in $M^2$
over a wider range of temperature. This inclusion is all the more necessary
for sum rules like the one for $\De m_{\rho}$, where the leading
operator contribution cancels largely with that of $\pi\pi$-continuum.

The higher dimension operators will, of course, complicate the evaluation of
the sum rules in that we have to know the temperature dependence of their
thermal averages. Also there is more proliferation of operators than what is
generally thought. The procedure in the literature \cite{Hatsuda}
of including dimension six quark
operators and excluding the Lorentz noninvariant gluon operators
\cite{foot3}, because of
the smallness of their coefficients by a factor of $\alpha_s(M^2)$, is not
justified. For, as we have seen, the quark and the gluon operators mix under
a change of scale, so that after renormalisation group inprovement, both the
coefficients are of the same order in $\alpha_s(M^2)$.

A way to proceed here is to write the entire set of sum rules by considering
two-point functions of not only the vector quark bilinear but also the
others, like the scalar, tensor, etc. All of these sum rules receive
contribution from a few resonances and the operators from the same set. The
simultaneous evaluation of all these sum rules is expected to provide a
self-consistent check on the thermal average of the operators. Further,
using quark bilinears of appropriate chiralities, one can get sum rules
without any of the gluon operators \cite{Mallik82}.
These sum rules should prove easier to evaluate and would also check the
saturation scheme in a simpler context.

\begin{center}
{\bf Acknowledgements}
\vspace{.3cm}
\end{center}

One of us (S.M.) is grateful to H. Leutwyler for helpful discussions 
when he was visiting the University of Bern, Switzerland. He also 
thanks M. Shapashnikov for a discussion of his own work. 

\renewcommand{\theequation}{A.\arabic{equation}}
\setcounter{equation}{0}
\begin{center}
{\bf Appendix}
\vspace{0.3cm}
\end{center}

Here we calculate the limit of the second integral in bracket in (3.17),
\be
A = {1\over {32\pi^2}}  \int^{\vq^2}_0 dq_0^2 e^{-(q_0^2 -\vq^2)/M^2}
\int_v^{\infty} dx x^2
 \{n((\vq x-q_0)/2)-n((\vq x+q_0)/2)\}
\ee
as $\vq\rightarrow 0$. It is convenient to change the integration variables
$\qz$ and $x$ to $\lm$ and u respectively by $\qz = \lm \vq$ and 
$\vq x = u$, getting
\[A= {1\over {32\pi^2}}  \int^1_0 d\lm^2 e^{-(\lm^2 -1)\vq^2/M^2}
\int_l^{\infty} du u^2{1\over \vq}
 \{n((\vq x-q_0)/2)-n((\vq x+q_0)/2)\} ,
\quad l= \sqrt{\vq^2 +{4m_{\pi}^2\over { 1-\lm^2}}} \]
As $\vq \rightarrow 0 $, it becomes
\ba
A &\rightarrow& -{1\over {16\pi^2}}\int ^1_0 d\lm^2 \lm 
\int^{\infty}_{2m_{\pi}\over{\sqrt{1-\lm^2}}}
du u^2 {dn(u/2)\over du} \nonumber \\
  &=& {1\over {48\pi^2}}
 \int ^{\infty}_{4m_{\pi}^2} ds v (3-v^2)n(\sqrt{s}/2) 
\ea
To get the second line we integrate by parts over $ u$ and interchange 
the order of integration in the remaining double integral.

The limit of the corresponding integral for $T_t$ is zero, because of the
presence of $q^2$ in the integrand.

\end{document}